# Giant Magneto-Exciton Coupling in 2D van der Waals CrSBr

Jia Shi,[1#] Dan Wang,[6#] Nai Jiang,[2,3#] Ziqian Xin,[1] Houzhi Zheng,[2,3] Chao Shen,[2,3*] Xinping Zhang,[1*] Xinfeng Liu[4,5*]

[1] Institute of Information Photonics Technology and School of Physics and Optoelectronics, Faculty of Science, Beijing University of Technology, Beijing 100124, China

[2]State Key Laboratory for Superlattices and Microstructures, Institute of Semiconductors, Chinese Academy of Sciences, Beijing 100083, China

[3]Center of Materials Science and Optoelectronics Engineering, University of Chinese Academy of Sciences, Beijing 100049, China

[4] CAS Key Laboratory of Standardization and Measurement for Nanotechnology, CAS Center for Excellence in Nanoscience, National Center for Nanoscience and Technology, Beijing 100190, China

[5] University of Chinese Academy of Sciences, Beijing, 100049, P. R. China

[6] State Key Laboratory of Low Dimensional Quantum Physics and Department of Physics, Tsinghua University, Beijing 100084, China

*Email address: liuxf@nanoctr.cn, zhangxinping@bjut.edu.cn, shenchao@semi.ac.cn

**Controlling magnetic order via external fields or heterostructures enables precise manipulation and tracking of spin and exciton information, facilitating the development of high-performance optical spin valves. However, the weak magneto-optical signals and instability of two dimensional (2D) antiferromagnetic (AFM) materials have hindered comprehensive studies on the complex coupling between magnetic order and excitons in bulk-like systems. Here, we leverage magneto-optical spectroscopy to reveal the impact of magnetic order on exciton-phonon coupling and exciton-magnetic order coupling which remains robust even under non-extreme temperature conditions (80 K) in thick layered CrSBr. A**

**0.425T in-plane magnetic field is sufficient to induce spin flipping and transition from AFM to ferromagnetic (FM) magnetic order in CrSBr, while magnetic circular dichroism (MCD) spectroscopy under an out-of-plane magnetic field provides direct insight into the complex spin canting behavior in thicker layers. Theoretical calculations reveal that the strong coupling between excitons and magnetic order, especially the 32 meV exciton energy shift during magnetic transitions, stems from the hybridization of Cr and S orbitals and the larger exciton wavefunction radius of higher-energy B excitons. These findings offer new opportunities and a solid foundation for future exploration of 2D AFM materials in magneto-optical sensors and quantum communication using excitons as spin carriers.**

Taking into account the spin or magnetic order as additional degree of freedom, the discovery of intrinsic 2D magnetic materials [1,2] unveils a host of novel physical phenomena, such as low-dimensional superconductivity,[3-5] giant magnetoresistance, [6] magnetic tunnel junctions,[7] *etc*. In 2D magnetic materials, excitons are crucial quasiparticles that shape the electronic structure.[8] Magnetic order dictates spin arrangement, affecting both the energy structure and excitonic properties.[9] Studying the coupling between excitons and other quasiparticles with magnetic order not only reveals new phases of matter, such as magnetic topology[10] and moiré magnetism,[11] but can also be applied to the development of spin-based technologies, like optical spin valves, spin-photonic devices, and ultra-sensitive magnetic field sensors.[12] Furthermore, due to their enhanced interactions between electron-spin-charge and lattice, 2D magnetic materials are highly tunable by external fields, making them ideal for design and integration into emerging heterostructures.[13-16] While 2D ferromagnetic materials have traditionally been favored for spintronic applications due to their straightforward magnetic properties and ease of control, their intrinsic stray fields pose challenges for device miniaturization.[17-20] 2D AFM materials offer a promising alternative, as they exhibit  produce negligible stray fields, making them more suitable for densely packed device arrays.[21]

However, the practical implementation of 2D AFM materials faces substantial obstacles. One of the primary challenges is the difficulty in efficiently detecting and manipulating the magnetic states. This limitation arises from their weak signal caused by the atomic layer thickness and nearly zero net magnetic moment, making traditional magnetic detection methods such as magnetic neutron scattering,[22]superconducting quantum interference device (SQUID)[23] and magneto-optical Kerr effect (MOKE)[24] inadequate. On the other hand, many 2D magnetic materials are unstable in air, prone to oxidation or degradation, which increases the complexity of their preparation and handling. [32-34]Moreover, 2D AFM materials often require extreme conditions such as ultra-low temperatures or strong magnetic fields between AFM and FM states like $NiPS_3$, which requires fields up to 6 T,[1,25,26]  further complicating their use in practical devices.

In our work, A-type CrSBr exhibits a significant coupling between magnetic order and excitons, demonstrating strong potential for applications in high-performance optical spin valves devices. Our study demonstrates several key findings regarding the strong coupling between excitons and magnetic order in thick layered CrSBr. Firstly, the magnetic phase transition from paramagnetic to antiferromagnetic order in CrSBr can be achieved by tuning temperature, and the coupling strength and modes between excitons and phonons are highly sensitive to these changes in magnetic order. Secondly, thick layered CrSBr allows for the transition from AFM to FM order under an external magnetic field. Along the easy magnetization axis, a modest field of 0.425T is sufficient to induce spin flipping in A and B excitons, with energy shifts of up to 20 meV. Moreover, applying an external magnetic field along the c-axis reveals the spin canting process via magnetic circular dichroism (MCD) spectroscopy. This technique enables direct observation of the more complex spin dynamics in thick CrSBr compared to thinner layers. Notably, even at moderate temperatures such as 80 K, the coupling between magnetic order and higher-energy B excitons remains remarkably strong, allowing for excitonic energy shifts of up to 32 meV during magnetic transitions. Theoretical calculations shed light on the underlying mechanism responsible for the large exciton–magnetic order coupling, attributing it to the larger wavefunction radius of the higher-energy B excitons and the hybridized orbital contributions from Cr and S atoms. These insights not only deepen our understanding but also pave the way for designing new materials or heterostructures with strong exciton–magnetic order coupling, suitable for applications in magneto-optical sensors and spin-switching devices.

We explore the thick-layer CrSBr (90 nm), which is an A-type van der Waals antiferromagnetic material, exhibiting antiferromagnetic order between layers with opposite spin directions and ferromagnetic order within layers with parallel spin alignment. [9,27-33] CrSBr has an orthorhombic crystal structure and belongs to the Pmmn space group. It exhibits an AA'-like stacking sequence along the c-axis, where each S-Cr-Br layer directly faces another S-Cr-Br layer, as illustrated in **Figure S1.** The calculated electronic structure of CrSBr with antiferromagnetic magnetic ordering is depicted in **Figure 1a.** The valence band maximum (VBM) and conduction band minimum (CBM) are both at the Γ point. Furthermore, there are two valence bands (denoted as $V_1$ and $V_2$) and two conduction bands ($C_1$ and $C_2$) near the band edge showing highly anisotropy along $\Gamma$-$Y$ and $\Gamma$-$X$. Note here the product symmetry of time reversal and spatial inversion in AFM state makes the band $V_1$, $V_2$, $C_1$ and $C_2$ all degenerate in spins. There are varied transition probabilities between these top two valence and bottom two conduction bands, which can be revealed by the calculated sum of the squares of the dipole transition matrix elements: $P^2$, along high-symmetry points as shown in **Figure S2a**. Based on the $P^2$ values, the transition between $V_1$ and $C_1$ ($V_2$ and $C_2$) is allowed, but the transition between $V_1$ and $C_2$ ($V_2$ and $C_1$) is dipole forbidden. This is similar to the monolayer case with inversion symmetry,[9] where $V_1$ and $C_2$ ($V_2$ and $C_1$) feature the same parity (**Figure S2b**) and thus the transitions between $V_1$ and $C_2$ ($V_2$ and $C_1$) are dipole-forbidden. Following those in TMDCs, we name here exciton transition from $V_1$ to $C_1$ at $\Gamma$ as A exciton which have been extensively studied[9] and $V_2$ to $C_2$ as B exciton. **Figure 1b** displays the photoluminescence (PL) spectra of CrSBr at room temperature and 10K, obtained under excitation with a 532 nm continuous laser at the same power. The PL intensity of A excitons at 1.3eV is one to two orders of magnitude higher than that of B excitons at 1.72 eV, with A excitons exhibiting a significantly narrower linewidth. At room temperature, the PL shape of A excitons approximates a symmetric Gaussian distribution, whereas the PL shape of B excitons is asymmetric. We hypothesize that this asymmetry arises from the flat-band nature of the B exciton transitions near the Γ point in the Brillouin zone. The flat-band leads to strong interactions between the excitons and lattice vibrations (phonons) during radiative recombination, resulting in the appearance of phonon sidebands adjacent to the main peak in the PL spectrum. Similar phonon sideband effects have been reported in other low-dimensional materials, such as monolayer $MoS_2$,[34,35] CdSe quantum dots,[36] and single-walled carbon nanotubes,[37]where exciton-phonon interactions significantly influence the spectral shape. **Figure 1c** shows the polarization-resolved PL peak distribution of A and B excitons at 10 K. As previously reported, the A exciton exhibits significant anisotropy, with maximum PL intensity along the b axis. Away from the *Γ* point, the conduction bands exhibit significant anisotropy, with dispersion along *Γ–Y* and nearly flat along *Γ–X,* consistent with the dipole-allowed interband transitions along the b axis but forbidden along a axis at the Γ point.[9] The B exciton also displays significant anisotropy, but its maximum PL intensity occurs at an angle approximately 15° off from the b-axis which is consistent with theory prediction.[38]

Additionally, we have performed temperature-dependent 2D contour differential reflectance spectroscopy, as illustrated in **Figure 1d**. The white lines enclosed within the yellow dashed frame represent the absorption peak positions of the A and B excitons, arranged from left to right, respectively. Both A and B excitons exhibit a significant red shift as the temperature decreases below 143 K. **Figure S3** and **Figure 1e** respectively illustrate the 2D contour plots of PL spectra of A and B excitons as a function of temperature, with intensity normalized for observing the variations in exciton peak positions and widths. The power dependent and normalized PL spectra peak shapes of A and B exciton are depicted in **Figure S4-5**. In the 2D contour PL spectra, a noticeable inflection point appears around 143 K for both the A and B excitons. Intriguingly, the A exciton exhibits a blue shift below 143 K, while the B exciton experiences a red shift. This transition in peak characteristics precisely accompanies the magnetic order transition from paramagnetic (PM) to AFM. This transition temperature ($T_N$=143 K) of thick layer CrSBr aligns well with the reported Neel temperature of CrSBr in the literature.[9] Here, we employ a Bose-Einstein oscillator model to analyze the underlying reasons for magnetic order related exciton-phonon coupling in CrSBr (**Figure 1f** and **Figure S6**). The model incorporates contributions from three optical modes, based on experimental data,[39-41] while excluding negligible acoustic mode contributions. Furthermore, given the magnetic order transition from PM to AFM around $T_N$, the model includes band perturbation terms related to magnetic ordering. The model equation is provided in Equation (1)

$$E_A(\mathrm{T,M}) = \mathrm{E(M)} + \frac{A_{LO1}}{M_{LO1}\omega_{LO1}}\left(\frac{1}{e^{\omega_{LO1}/KT}-1}+\frac{1}{2}\right)+\frac{A_{LO2}}{M_{LO2}\omega_{LO2}}\left(\frac{1}{e^{\omega_{LO2}/KT}-1}+\frac{1}{2}\right)+\frac{A_{LO3}}{M_{LO3}\omega_{LO3}}\left(\frac{1}{e^{\omega_{LO3}/KT}-1}+\frac{1}{2}\right) \quad (1)$$

Where $E(M)$ describes the contribution from the magnetic order to the change of bandgap. $A_{LO1}$, $A_{LO2}$ and $A_{LO3}$ represent the relative weight of oscillators (or the electron–phonon coupling coefficients), $\omega_{LO1}$ , $\omega_{LO2}$ and $\omega_{LO3}$ are the optical phonon energies $A_g^1$ (118cm$^{-1}$), $A_g^2$ (244cm$^{-1}$), and $A_g^3$ (344cm$^{-1}$), corresponding to Br, Cr, and S atoms, respectively.[39] $M_{LO1}$, $M_{LO2}$ and $M_{LO3}$ represent the atomic masses of the oscillators. **Table 1** shows the fitting results. It clearly indicates that the contributions of $E(M)$ are significantly affected by magnetic order, with shifts reach 30 to 40 meV. This highlights the critical role of magnetic order in the exciton-related electronic structure of CrSBr. The red and blue shifts of the A and B exciton PL peaks are dominated by the shifts in electronic energy levels during the transition from PM to AFM states. Additionally, model parameters reveal that as the exciton-phonon coupling shifts from involving three optical phonons to primarily the optical phonons associated with S and Cr atoms with changing from PM to AFM states for A exciton. In terms of B exciton, electron-phonon coupling shifts from being dominated by Cr-related phonons in the PM state to S-related phonons in the AFM state. This shift is likely due to changes in the electronic environment and bonding characteristics influenced by magnetic ordering. Additionally, the coupling coefficients for S and Cr- related phonons change sign during the magnetic state transition, while Br -related phonon coupling remains stable, indicating that Br atoms are less affected by magnetic order.

Building on our understanding of magnetic order related exciton-phonon coupling in CrSBr from PM to AFM transitions, we applied in-plane and out-of-plane magnetic fields to analyze the coupling differences in the electronic structure for A and B excitons under AFM and induced in-plane and out-of-plane FM states. The PL peak positions of the A and B excitons in CrSBr redshift and their intensity significantly increases when an external magnetic field along b axis (easy axis) reaches a critical value of 0.425 T as shown in **Figure 2a-b**. The critical magnetic field for bilayer and four-layer CrSBr is 0.13 T and 0.33 T, respectively, while 90nm thick CrSBr requires only 0.425 T to realize the spin flipping. This is highly encouraging compared to other materials,[9] suggesting that CrSBr is easily tunable by external magnetic fields and highly sensitive to magnetic order from thin layers to bulk. The A exciton transition energy changes by 14 meV from AFM to in-plane FM state, and the B exciton changes by 20 meV. Additionally, the PL intensity of the B exciton significantly increases with magnetic order change, indicating stronger coupling between magnetic order and electronic structure for B excitons. Since the application of an in-plane magnetic field leads to a more uniform spin alignment across the layers, this uniformity reduces spin-related scattering events, thereby narrowing the exciton linewidth.

When the external magnetic field along b axis reaches saturation fields, CrSBr transitions from AFM state to in-plane FM state, resulting in abrupt PL peak shifts of A and B excitons. This is due to spin flip-induced changes in interlayer magnetic ordering. Conversely, when the magnetic field is applied along the c axis (hard axis), the spin canting process leads to a transition from AFM to interlayer FM state. To investigate this, we employed energy-resolved MCD spectrum to monitor and analyze the electronic structure modulated by spin canting process. Since the MCD signal originates from the Zeeman splitting of excitons under an applied magnetic field perpendicular to the sample surface, breaking valley degeneracy,[42,43] the MCD spectra can provide a direct reflection of the magnetic field's modulation of the electronic structure as shown in **Figure 3a**. As the external magnetic field along the c-axis increases, the in-plane oriented spins gradually cant, resulting in a net magnetic moment along the c-axis. This process is accompanied by a minor component of uncompensated in-plane magnetic moment. **Figure 3b-c** shows the MCD spectra of CrSBr (90 nm) at 80 K, with the applied magnetic field ranging from -6T to 6T. Due to the difference in the real part of the refractive index of CrSBr for left and right circularly polarized light under a magnetic field, the MCD can be considered as the differential spectrum of the reflectance spectra. At a given magnetic field, the MCD can be expressed as:

$$MCD = \frac{\Delta r}{(r_L + r_R)/2} \approx \frac{1}{r}\frac{dr}{dn}\Delta n \qquad (3)$$

Here, $r_R$ and $r_L$ represent the reflectance of right- and left-circularly polarized light, $\Delta n$ is the is the difference in the refractive index of CrSBr for right-circularly polarized light and left-circularly polarized light. The average of $r_R$ and $r_L$ can be considered as the intensity of unpolarized light.

**Figure 3b** presents a 2D contour plot of the MCD spectra, akin to the first derivative of reflectance spectra. The white lines indicate zero-crossing points of the excitonic energy levels, with the MCD signal reversing as the magnetic field polarity changes. The plot clearly shows distinct behaviors of various excitonic states with increasing magnetic field strength. Notably, the energy levels of the A and B excitons exhibit a redshift with increasing magnetic field, consistent with our previous conclusions. To eliminate the influence of laser intensity on reflectance, we calculate the difference in reflectance between $r_R$ and $r_L$, divided by their average reflectance, which gives the value of MCD and the intensity of MCD is proportional to the magnitude of the magnetic field as shown in **Figure 3c**. From the MCD spectrum, the energy levels corresponding to the A and B excitons are identified as 1.35 eV and 1.70 eV, respectively, which are consistent with the values obtained from our previous reflectance spectra measurements. Additionally, the MCD spectra reveal excitonic energy levels at 1.51 eV, 1.60 eV, and 1.80 eV, corresponding to higher-energy excitonic states such as "*2s*" and "*2p*", *etc* reported in previous theoretical calculations literature[38]. The overlap of these higher-energy excitonic states with the "1s" states of A and B excitons in the MCD spectra results in asymmetry peak shape around the zero-crossing point. This overlap prevents accurate fitting of the Zeeman splitting induced by the external magnetic field. The field-dependent MCD spectra at 1.73 eV, shown in **Figure 3c**, were used to derive the magnetization loop (**Figure 3d**), revealing a saturation magnetic field of approximately 4T along the out-of-plane hard axis (c-axis). The differential reflectance spectra and MCD spectra in **Figure 3e-f**, obtained by scanning the magnetic field along c-axis, validate the spin canting process. When the magnetic field is applied along the hard axis, spin canting induces continuous changes in interlayer magnetic order and electronic structure. This results in a continuous redshift of the exciton peaks in the FM state compared to the AFM state. Notably, the A exciton shifts by ~12 meV under the saturation field, while the B exciton exhibits a significant redshift of ~32 meV. These findings indicate that the dramatic changes in excitonic properties are due to adjustments in interlayer magnetic order. The B exciton in CrSBr is more sensitive to interlayer electronic coupling, making it more responsive to the out-of-plane hard axis magnetic field. Additionally, within the magnetic field range of -1 to 1T, the excitonic energy levels shift rapidly during the spin canting process, but the redshift slows down beyond 1T. This may be due to a critical state within the -1 to 1T range, where an in-plane AFM and interlayer FM mixed state exists, affecting the electronic levels differently along the c-axis and b-axis. The asymmetry in energy levels under positive and negative magnetic fields could be due to the interlayer AFM order with in-plane spin alignment in multilayer CrSBr at 80 K. With a thickness of 90 nm, odd-numbered layers may have uncompensated magnetic moments, resulting in a residual ferrimagnetic state[44]. As the external magnetic field along the c-axis increases, the in-plane spins cant, producing a net magnetic moment along the c-axis and a minor component of uncompensated in-plane magnetic moment. The magnetic field dependent PL and MCD spectra reveal that the electronic structure of thick layered CrSBr is highly sensitive to magnetic order, whether subjected to in-plane or out-of-plane magnetic fields. The higher-energy B exciton shows greater sensitivity to magnetic ordering than the A exciton. Under the same saturation magnetic field, the coupling between magnetic order and electronic structure results in a redshift in the energy levels of the B exciton that is approximately three times greater than that of the A exciton.

**Theoretical Insights into Magnetic and Excitonic Coupling**

To investigate the theoretical underpinnings of the differing optical behaviors of A and B excitons, we focus on the orbital-resolved electronic structure of CrSBr, particularly from the Γ point to the X point in the Brillouin zone, corresponding to the transition levels of A and B excitons. **Figure 4a, c** and **Figure S8a** illustrate the orbital contributions of Cr, S, and Br atoms. The valence bands $V_1$ and $V_2$ primarily arise from Cr: $d_{yz}$ , S: $p_y$ , and Br: $p_y$ orbitals, with $V_2$ exhibiting a greater S: $p_y$ contribution than $V_1$. The conduction bands $C_1$ and $C_2$ mainly derive from Cr: $d_{x^2-y^2}$ and $d_{z^2}$ orbitals, with $C_2$ also incorporating contributions from S: $p_z$ orbitals. This indicates that the B exciton ($C_2$-$V_2$) involves more substantial S: $p_z$ orbitals, rendering B excitons more susceptible to interlayer coupling, which is consistent with the significant red shift when applying an out-of-plane magnetic field as discussed above. This hypothesis is further corroborated by the partial charge plots for $C_1$, $C_2$, $V_1$, and $V_2$ near the Γ point, as shown in **Figure S7c**. Additionally, we calculated the electronic structure of monolayer CrSBr **(Figure S9)** and found that its distribution closely matches that of the bulk structure. Due to computational limitations in calculating the A and B exciton wavefunctions for the bulk structure using Yambo: GW-BSE, we opted to calculate the wavefunctions for A and B excitons in monolayer CrSBr instead. The results are shown in **Figure 4e-f**, where the red spheres indicate the fixed positions of the holes. In the AFM state, both A and B excitons are intralayer excitons (**Figure S10**). However, the Bohr radius of the B exciton is 2-3 times larger than that of the A exciton, making B excitons more susceptible to changes in the lattice, magnetic order and interlayer coupling. Given the complexity of spin canting in bulk-like CrSBr, most studies have focused on bilayer and few-layer systems.[9] However, this work combines theoretical calculations and MCD experiments to provide a clear observation of the spin canting process in thick CrSBr. We theoretically analyzed the spin canting behavior of A and B excitons induced by applying a magnetic field along the hard magnetization axis (c axis). **Figure 4g** also schematically illustrates the tilting of Cr atom spins in multilayer CrSBr under an external magnetic field along the c-axis. The angle $\theta$ represents the tilt between the spin direction and the c-axis, transitioning from *90°* in the AFM state to *0º* in the out-of-plane FM state as the magnetic field increases. We calculated the electronic band structure of bulk CrSBr with $\theta$ varying from *90º* to *0º* in *15°* increments, as shown in **Figure S15**. From the initial AFM state to the final out-of-plane FM state, $C_1$, $C_2$, $V_1$, and $V_2$ split as the angle decreases, causing a redshift in the transition energies of A and B excitons. For the interlayer AFM order, the spin-up and spin-down electrons in each band are degenerate in energy but localized at different layers, since their interlayer hybridization is suppressed. In the contrast, when spin canting induces continuous changes to the interlayer magnetic order, the electrons in the adjacent layers can resonantly couple with each other, leading to band splitting of the $C_1$, $C_2$, $V_1$, and $V_2$ relative to the AFM bilayer. **Figure 4h** shows the allowed excitonic transition energies corresponding to the excitonic states permitted by the selection rules during the AFM-to-FM spin canting process. During spin canting process the most direct effect is a redshift of the optically bright A and B excitons from the zero-field AFM state. Comparing **Figure 4b** and **4d**, the band split energies for $C_1$, $C_2$, $V_1$, and $V_2$ are roughly in the same magnitude. However, when $\theta$ gradual decrease, the split $C_1^2$ and $C_2^1$ band are getting closer and hybridization happens between them as shown in **Figure 4c-d** and **Figure S16** ($C_1^2$ gains some S: $P_z$ contributions) in FM state. Eventually, optical transition from $V_1^1$ to $C_1^2$ are allowed and mixed with $V_1^1$ to $C_2^1$, which qualitatively explains the more promote redshift of B exciton across the spin canting process., based on orbital contributions and allowed transition rules. Although the bandgap values corresponding to A and B excitons obtained using the PBE method cannot be directly compared with experimental results, the trend and relative magnitude of these shifts are consistent.

Finally, we compared the typical saturation reversal magnetic fields of other reported 2D magnetic materials (**Table S2**). Thin-layer materials like $CrI_3$ and $Cr_2Ge_2Te_6$ require low saturation fields but degrade easily in air. Thicker materials such as $Fe_3GeTe_2$ and $FePS_3$ require fields exceeding 6 T and even 35 T. In contrast, the quasi-2D CrSBr material reported here remains stable at room temperature and requires only 0.425 T in-plane and 4T out-of-plane to alter magnetic order. It exhibits strong magneto-exciton coupling with significant B-exciton energy tuning (32 meV at 80 K). These properties make CrSBr a promising, field-tunable antiferromagnetic material for next-generation spintronic and magneto-optical devices.

## Conclusion

Our comprehensive study on the intrinsic coupling between magnetic order and excitonic states in CrSBr reveals significant findings. The B exciton, compared to the A exciton, shows greater sensitivity to magnetic order changes, particularly during spin canting transitions from AFM to FM states. This is evidenced by substantial redshifts in the excitonic transition energies, confirmed through both theoretical calculations and magneto-optical experiments. These results underscore the potential of CrSBr as a tunable 2D magnetic semiconductor for applications in spintronics and optoelectronics. The clear observation of spin canting effects in thick-layered CrSBr further enhances our understanding of magnetic order interactions in layered materials, paving the way for future explorations and technological advancements.

## Methods

***Optical characterizations:*** The magneto-PL measurements were carried out in a Montana system at a sample temperature of 10 K with the excitation of the 532-nm laser. The emission light was collected by a Horiba iHR320 system and the applied magnetic field was along the b direction of the sample surface. The MCD measurement was conducted using a wavelength-tunable laser derived from a supercontinuum white light source equipped with a monochromator (Horiba Jobin-Yvon iHR320). Detection of the reflected light from the sample was achieved using a Si photodetector. Reflectance and MCD signals were measured with two lock-in amplifiers, referencing a 177 Hz chopper and a 50 kHz photoelastic modulator, respectively. And the applied out-of-plane magnetic fields along the incident direction of the excitation laser were provided by a superconducting magnet.

**First-principles calculations:** The DFT calculations were carried out using the projector augmented wave method[45] as implemented in VASP package.[46] The generalized gradient approximation (GGA) functional of Perdew−Burke−Ernzerhof (PBE)[47,48] was adopted to characterize the electron exchange and correlation effects with a kinetic energy cutoff of 500 eV. The convergence thresholds for electronic and ionic relaxations were chosen to be $1.0 \times 10^{-7}$ eV and 0.001 eV $Å^{-1}$. The Brillouin zones for the bulk CrSBr were sampled by $10 \times 7 \times 4$ Γ-centered k-point grids. The SOC effect and dispersion corrections within the DFT-D3 formalism to account for the vdW interactions[49] were included in all our calculations. For the bilayer CrSBr, a single-shot GW ($G_0W_0$)[50] approach was employed to calculate the quasiparticle band structure and using the Yambo code.[51,52] The dielectric functions were obtained by solving the Bethe−Salpeter equation (BSE)[53] on the basis of the $G_0W_0$ calculations. The exciton properties were captured using GW-BSE method on a 16×12×1 k-grid. A truncated Coulomb interaction was employed in both the $G_0W_0$ and BSE calculations to prevent the interlayer screening interaction between the periodic layers along the out-of-plane direction. Eight valence bands and eight conduction bands were included during the BSE calculations.

## Associated content

*Supplementary Information

The Supplementary Information is available.
Figure S1-S16 and calculation as described in the text.

**Author information**

Corresponding Authors
*E-mail: liuxf@nanoctr.cn, zhangxinping@bjut.edu.cn, shenchao@semi.ac.cn

**Author contributions:** J.S., X.L., X.Z. and C.S. lead the project. J.S., N.J., and Z.X. conducted the magneto-optical experiments. S.Z. provided the samples. X.L., C.S., X.Z. led the experimental design and analysis. D.W. was responsible for the theoretical calculations. J.S. and X.L. wrote the manuscript. All authors discussed the results and revised the manuscript. J.S., D.W., and N.J. contributed equally to this work.

**Acknowledgments**

**Funding:** This work was supported by the National Natural Science Foundation of China (62305012), the Beijing Postdoctoral Research Foundation (2023-zz-60), and the China Postdoctoral Science Foundation (2023M730155). We also thank the support from National Key Research and Development Program of China (2023YFA1507002), National Science Foundation for Distinguished Young Scholars of China (22325301). We are deeply grateful to Professor Zhou Shengqiang for providing the CrSBr samples and engaging in discussions with us.

**Notes**

The authors declare no competing financial interest.

**References:**

1. Gong C*, et al.* Discovery of intrinsic ferromagnetism in two-dimensional van der Waals crystals. *Nature* **546**, 265-269 (2017).
2. Huang B*, et al.* Layer-dependent ferromagnetism in a van der Waals crystal down to the monolayer limit. *Nature* **546**, 270-273 (2017).
3. Sohn E*, et al.* An unusual continuous paramagnetic-limited superconducting phase transition in 2D $NbSe_2$. *Nat. Mater.* **17**, 504-508 (2018).
4. Kezilebieke S*, et al.* Topological superconductivity in a van der Waals heterostructure. *Nature* **588**, 424-428 (2020).
5. Machida T*, et al.* Zero-energy vortex bound state in the superconducting topological surface state of Fe(Se,Te). *Nat. Mater.* **18**, 811-815 (2019).
6. Telford EJ*, et al.* Layered Antiferromagnetism Induces Large Negative Magnetoresistance in the van der Waals Semiconductor CrSBr. *Adv. Mater.* **32**, 2003240 (2020).
7. Chen Y*, et al.* Twist-assisted all-antiferromagnetic tunnel junction in the atomic limit. *Nature* **632**, 1045-1051 (2024).
8. Du L. New excitons in multilayer 2D materials. *Nat. Rev. Phys.* **6**, 157-159 (2024).
9. Wilson NP*, et al.* Interlayer electronic coupling on demand in a 2D magnetic semiconductor. *Nat. Mater.* **20**, 1657-1662 (2021).
10. Abuawwad N, dos Santos Dias M, Abusara H, Lounis S. Electrical engineering of topological magnetism in two-dimensional heterobilayers. *npj Spintronics* **2**, 10 (2024).
11. Song T*, et al.* Direct visualization of magnetic domains and moiré magnetism in twisted 2D magnets. *Science* **374**, 1140-1144 (2021).
12. Burch KS, Mandrus D, Park J-G. Magnetism in two-dimensional van der Waals materials. *Nature* **563**, 47-52 (2018).
13. Mak KF, Shan J, Ralph DC. Probing and controlling magnetic states in 2D layered magnetic materials. *Nat. Rev. Phys.* **1**, 646-661 (2019).
14. Tabataba-Vakili F*, et al.* Doping-control of excitons and magnetism in few-layer CrSBr. *Nat. Commun.* **15**, 4735 (2024).
15. Gibertini M, Koperski M, Morpurgo AF, Novoselov KS. Magnetic 2D materials and heterostructures. *Nat. Nanotechnol.* **14**, 408-419 (2019).
16. Long F*, et al.* Ferromagnetic Interlayer Coupling in CrSBr Crystals Irradiated by Ions. *Nano Lett.* **23**, 8468-8473 (2023).
17. McGuire MA, Dixit H, Cooper VR, Sales BC. Coupling of crystal structure and magnetism in the layered, ferromagnetic insulator $CrI_3$ *Chemistry of Materials* **27**, (2014).
18. Zhang W-B, Qu Q, Zhu P, Lam C-H. Robust intrinsic ferromagnetism and half semiconductivity in stable two-dimensional single-layer chromium trihalides. *J. Mater. Chem. C* **3**, 12457-12468 (2015).
19. Ningrum VP*, et al.* Recent Advances in Two-Dimensional Magnets: Physics and Devices towards Spintronic Applications. *Research* **2020**.
20. Lei Z*, et al.* Manipulation of ferromagnetism in intrinsic two-dimensional magnetic and nonmagnetic materials. *Matter* **5**, 4212-4273 (2022).

21. Jungwirth T, Marti X, Wadley P, Wunderlich J. Antiferromagnetic spintronics. *Nat. Nanotechnol.* **11**, 231-241 (2016).
22. Schins AG, Arts AFM, de Wijn HW, Nielsen M. Neutron scattering study of a two-dimensional ferromagnet with competing anisotropies. *J. Magn. Magn. Mater.* **104-107**, 931-932 (1992).
23. Farrar LS, Nevill A, Lim ZJ, Balakrishnan G, Dale S, Bending SJ. Superconducting Quantum Interference in Twisted van der Waals Heterostructures. *Nano Letter.* **21**, 6725-6731 (2021).
24. Wu H*, et al.* Strong intrinsic room-temperature ferromagnetism in freestanding non-van der Waals ultrathin 2D crystals. *Nat. Commun.* **12**, 5688 (2021).
25. Wang QH*, et al.* The magnetic genome of two-dimensional van der waals materials. *ACS Nano* **16**, 6960-7079 (2022).
26. Kang L*, et al.* Phase-controllable growth of ultrathin 2D magnetic FeTe crystals. *Nat. Commun.* **11**, 3729 (2020).
27. Ziebel ME, Feuer ML, Cox J, Zhu X, Dean CR, Roy X. CrSBr: An Air-Stable, Two-Dimensional Magnetic Semiconductor. *Nano Lett.* **24**, 4319-4329 (2024).
28. Klein J*, et al.* The bulk van der Waals layered magnet CrSBr is a quasi-1D material. *ACS Nano* **17**, 5316-5328 (2023).
29. López-Paz SA*, et al.* Dynamic magnetic crossover at the origin of the hidden-order in van der Waals antiferromagnet CrSBr. *Nat. Commun.* **13**, 4745 (2022).
30. Moro F*, et al.* Revealing 2D magnetism in a bulk CrSBr single crystal by electron spin resonance. *Adv. Funct. Mater.* **32**, 2207044 (2022).
31. Meineke C*, et al.* Ultrafast exciton dynamics in the atomically thin van der Waals magnet CrSBr. *Nano Lett.* **24**, 4101-4107 (2024).
32. Rizzo DJ*, et al.* Visualizing atomically layered magnetism in CrSBr. *Adv. Mater.* **34**, 2201000 (2022).
33. Long F*, et al.* Intrinsic magnetic properties of the layered antiferromagnet CrSBr. *Appl. Phys. Lett.* **123**, 222401 (2023).
34. Shree S*, et al.* Observation of exciton-phonon coupling in $MoSe_2$ monolayers. *Phys. Rev. B* **98**, 035302 (2018).
35. He K*, et al.* Tightly bound excitons in monolayer $WSe_2$. *Phys. Rev. Lett.* **113**, 026803 (2014).
36. Norris DJ, Sacra A, Murray CB, Bawendi MG. Measurement of the size dependent hole spectrum in CdSe quantum dots. *Phys. Rev. Lett.* **72**, 2612-2615 (1994).
37. Shreve AP*, et al.* Determination of exciton-phonon coupling elements in single-walled carbon nanotubes by raman overtone analysis. *Phys. Rev. Lett.* **98**, 037405 (2007).
38. Qian T-X, Zhou J, Cai T-Y, Ju S. Anisotropic electron-hole excitation and large linear dichroism in the two-dimensional ferromagnet CrSBr with in-plane magnetization. *Phys. Rev. Res.* **5**, 033143 (2023).
39. Lin K*, et al.* Strong exciton–phonon coupling as a fingerprint of magnetic ordering in van der Waals layered CrSBr. *ACS Nano* **18**, 2898-2905 (2024).

40. Linhart WM*, et al.* Optical markers of magnetic phase transition in CrSBr. *J. Mater. Chem. C* **11**, 8423-8430 (2023).
41. Xu X, Wang X, Chang P, Chen X, Guan L, Tao J. Strong Spin-Phonon Coupling in Two-Dimensional Magnetic Semiconductor CrSBr. *J. Mater. Chem. C* **126**, 10574-10583 (2022).
42. Xia M*, et al.* Magnetic circular dichroism study of electronic transition in metal $Fe_3GeTe_2$. *J. Mater. Chem. C* **126**, 8152-8157 (2022).
43. Wu YJ*, et al.* Valley Zeeman splitting of monolayer $MoS_2$ probed by low-field magnetic circular dichroism spectroscopy at room temperature. *Appl. Phys. Lett.* **112**, (2018).
44. Ye C*, et al.* Layer-dependent interlayer antiferromagnetic spin reorientation in air-stable semiconductor CrSBr. *ACS Nano* **16**, 11876-11883 (2022).
45. Kresse G, Joubert D. From ultrasoft pseudopotentials to the projector augmented-wave method. *Phys. Rev. B* **59**, 1758-1775 (1999).
46. Kresse G, Furthmüller J. Efficiency of ab-initio total energy calculations for metals and semiconductors using a plane-wave basis set. *Comput. Mater. Sci.* **6**, 15-50 (1996).
47. Perdew JP, Burke K, Ernzerhof M. Generalized Gradient Approximation Made Simple. *Phys. Rev. Lett.* **78**, 1396-1396 (1997).
48. Zhang Y, Yang W. Comment on ``Generalized Gradient Approximation Made Simple''. *Phys. Rev. Lett.* **80**, 890-890 (1998).
49. Grimme S. Semiempirical GGA-type density functional constructed with a long-range dispersion correction. *J. Comput. Chem.* **27**, 1787-1799 (2006).
50. Hybertsen MS, Louie SG. Electron correlation in semiconductors and insulators: Band gaps and quasiparticle energies. *Phys. Rev. B* **34**, 5390-5413 (1986).
51. Marini A, Hogan C, Grüning M, Varsano D. yambo: An ab initio tool for excited state calculations. *Comput. Phys. Commun.* **180**, 1392-1403 (2009).
52. Sangalli D*, et al.* Many-body perturbation theory calculations using the yambo code. *J. Condens. Matter Phys.* **31**, 325902 (2019).
53. Rohlfing M, Louie SG. Electron-hole excitations and optical spectra from first principles. *Phys. Rev. B* **62**, 4927-4944 (2000).

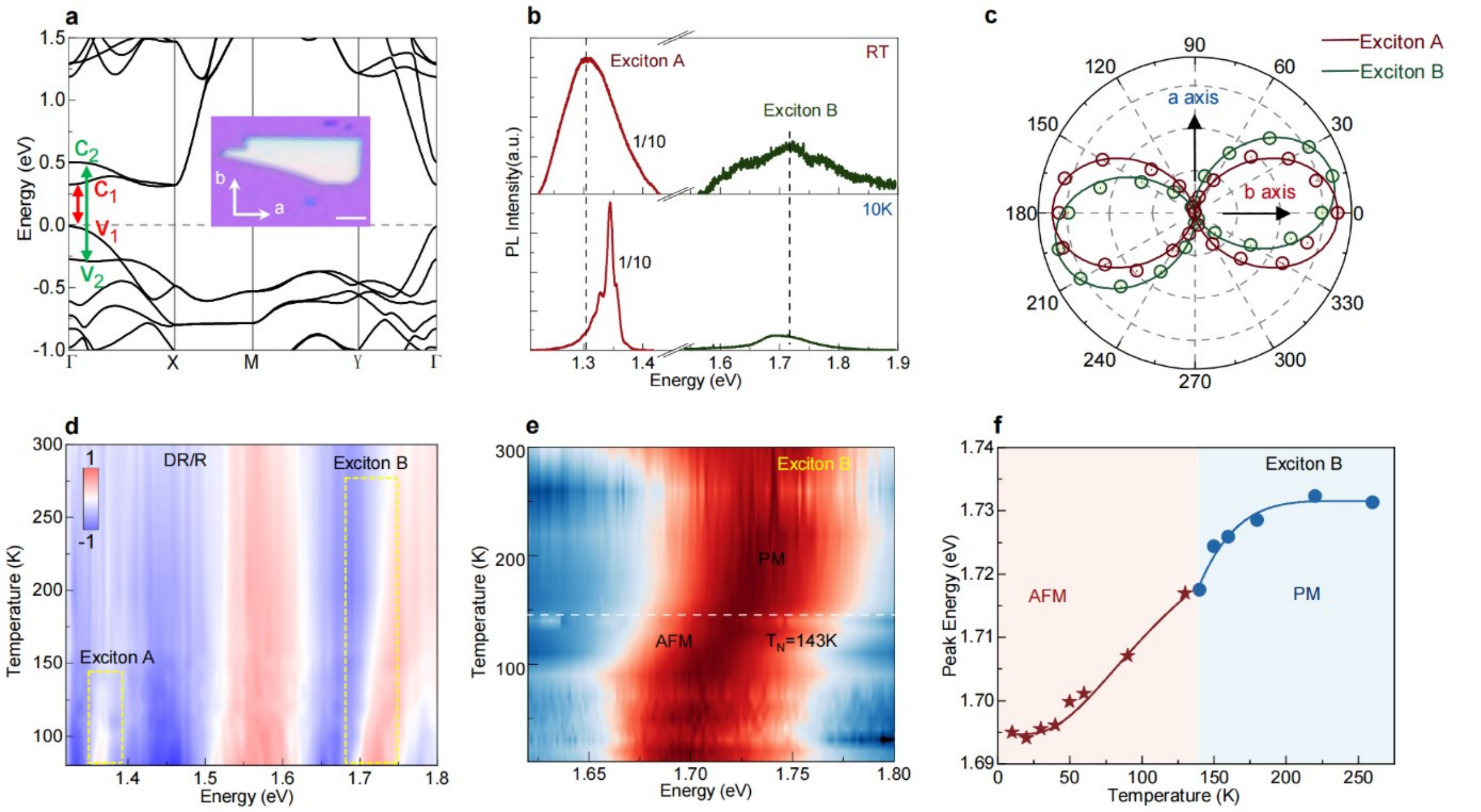

**Figure 1| Schematic Diagram of Transitions and Fundamental Optical Properties of A and B Excitons**. (**a**) The band structure of CrSBr, where the transition $C_1$-$V_1$ corresponds to A exciton and $C_2$-$V_2$ corresponds to B exciton.( The valence band maximum (VBM) is selected as the zero energy level here.) Inset: An optical photograph of CrSBr with a thickness of 90 nm. (**b**) Room temperature and 10 K PL spectra of CrSBr. (**c**) Polarization-dependent anisotropic PL spectra of A and B excitons measured at 10 K. (**d**) Temperature-dependent micro-area differential reflectance spectra of CrSBr. (**e**) Temperature-dependent PL spectra of B excitons, with normalized intensity to observe shifts in peak position and bandwidth. (**f**) Temperature-dependent PL peak positions of B excitons in CrSBr, fitted using the Bose-Einstein condensation model. Variations in exciton-phonon coupling strength are observed between PM and AFM states

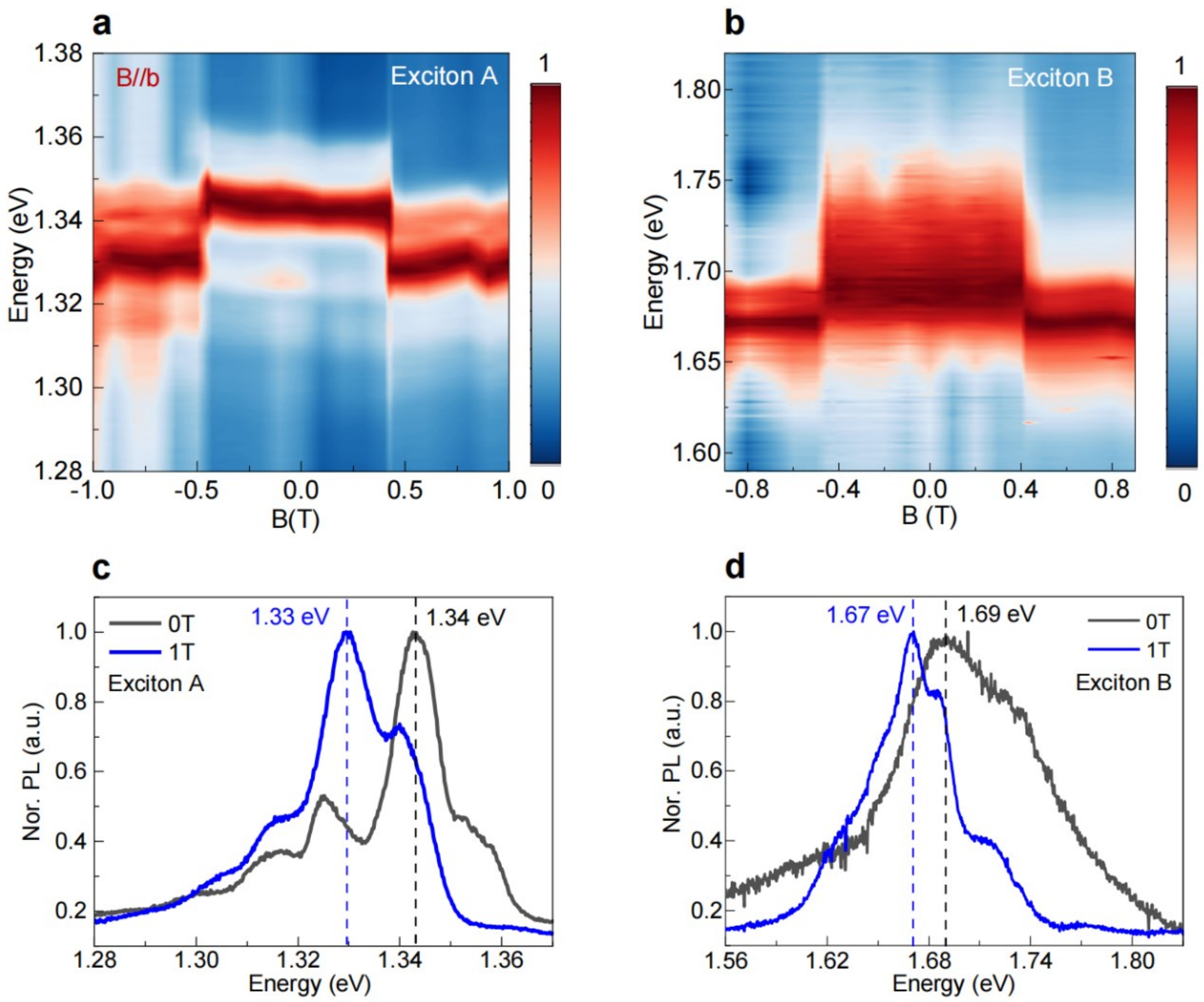

**Figure 2| Magnetic Field-Induced Electronic Structure Transformation from AFM to FM.** (**a**)-(**b**) 2D contour plot of PL spectra of A and B exciton in CrSBr under varying magnetic fields along the b axis. (**c**)-(**d**) Normalized PL spectra of A and B excitons in CrSBr at 0*T* and 1*T*, the intensity has been normalized to facilitate the observation of peak shifts.

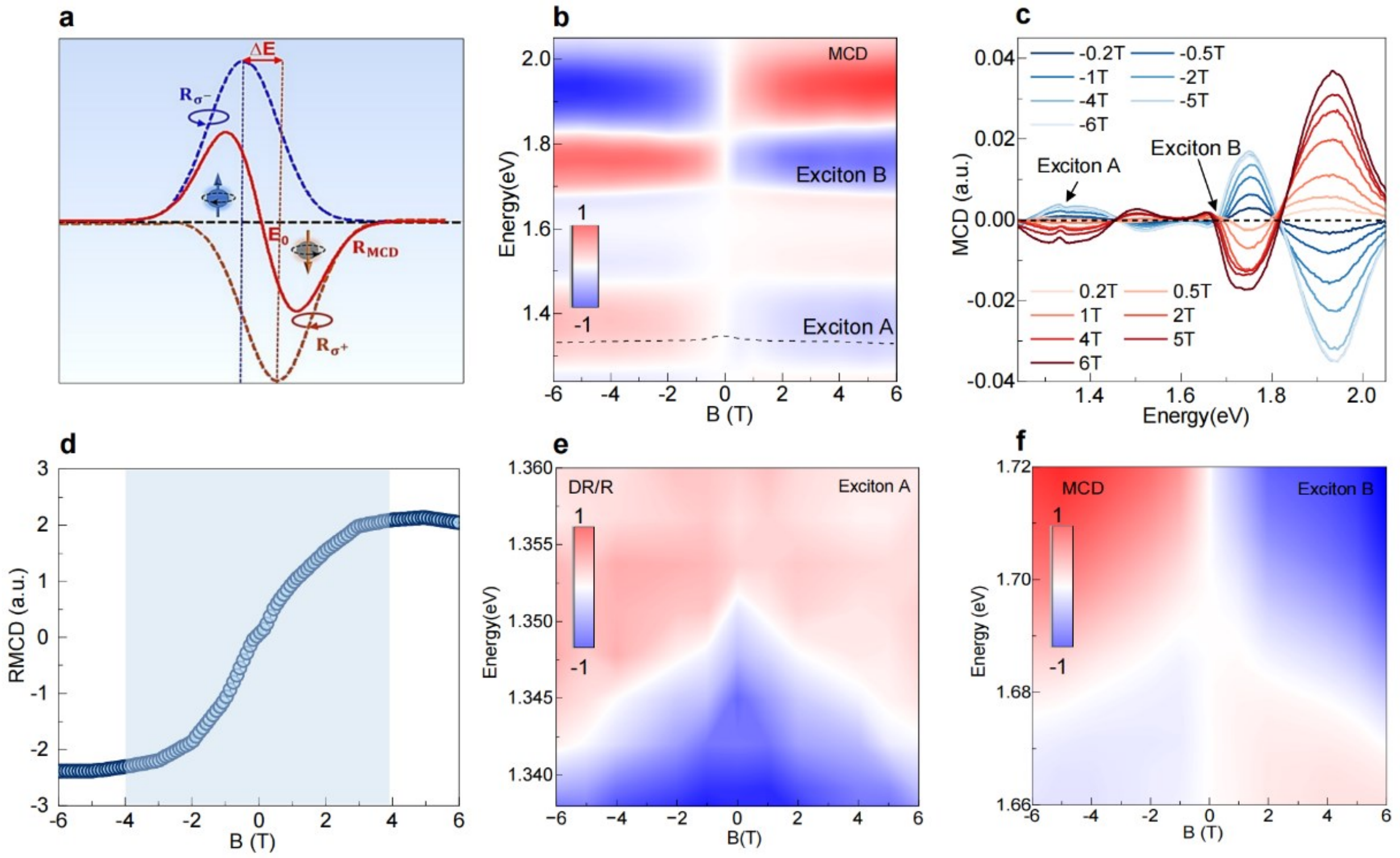

**Figure3| Magnetic Field-Induced Electronic Structure Transformation from in-plane AFM to out of plane FM.** (**a**) Illustration of the MCD principle. (**b**) 2D contour plot of MCD spectra, analogous to the first derivative of reflectance spectra. White lines indicate zero-crossing points of excitonic energy levels, with MCD signals reversing polarity as the magnetic field changes. (**c**) MCD spectra of multilayer CrSBr at 80 K under varying magnetic fields from -6 T to 6 T. (**d**) Magnetization loop along the c-axis derived from the MCD spectra at 1.73 eV, showing a saturation magnetic field of approximately 4 T. (**e**) Magnetic field dependent micro-area differential reflectance spectra of A exciton. (**f**) MCD spectra of B exciton at various magnetic fields along the hard c-axis, validating the spin canting process. Continuous changes in interlayer magnetic order and electronic structure are observed.

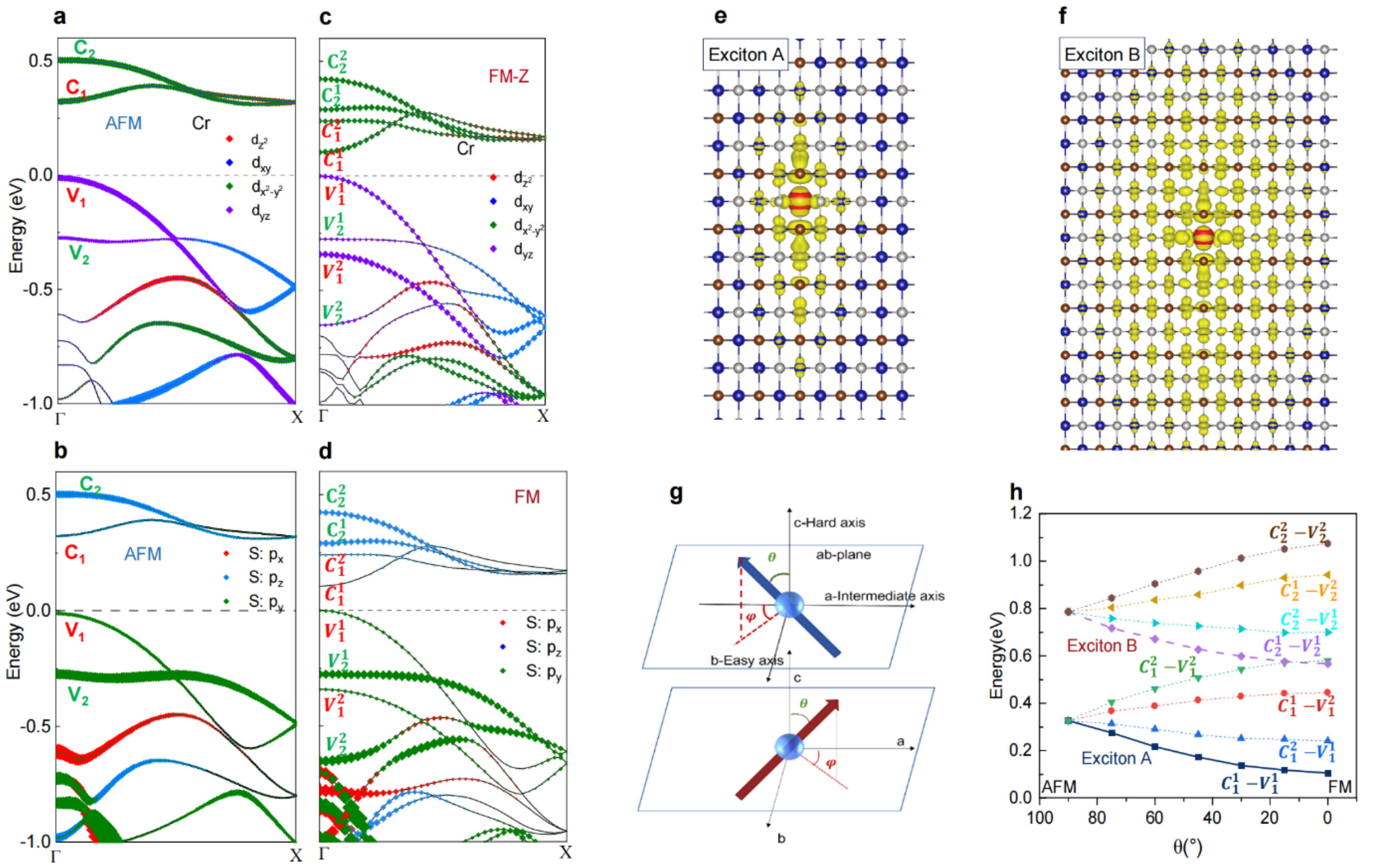

**Figure4| Theoretical calculation of Electronic Structure in bulk CrSBr and Excitonic States in Bilayer CrSBr.** (**a**)-(**d**) The orbital-decomposed band structure of bulk CrSBr in the AFM (left), in-plane FM (right) state. Top and bottom panels correspond to the orbital contributions of Cr and S atoms, respectively. (**e**)-(**f**) Real-space distributions of the exciton wave function for the A and B exciton, respectively. Here the hole is fixed at Cr atom. (**g**) schematic diagram illustrates the spin canting of Cr atoms within different layers of CrSBr induced by an external magnetic field along the c-axis. (**h**) The calculated bandgap at $\Gamma$ correlates to allowed excitonic transition energies permitted by the selection rules as a function of the spin canting angle $\theta$ ($\theta = 90^{\circ}, 75^{\circ}, 45^{\circ}, 30^{\circ}, 15^{\circ}$ and $0^{\circ}$).

**Table 1. Parameters of A and B excitons derived via fitting the PL peak energy temperature dependence using a Bose–Einstein mode**

| Exciton Type | $E_M$(eV) | $A_{LO1}$(Br) | $A_{LO2}$(Cr) | $A_{LO3}$(S) | $\omega_{opt1}$ (meV) | $\omega_{opt2}$ | $\omega_{opt3}$ |
|---|---|---|---|---|---|---|---|
| **PM A exction** | 1.34±0.04 | -0.96±0.10 | 3.00±0.20 | -1.80 ±0.04 | 14 | 30 | 43 |
| **AFM A exciton** | 1.37±0.01 | -0.06±0.02 | -0.42±0.01 | 0.48±0.01 | 14 | 30 | 43 |
| **PM B exction** | 1.74±0.01 | 1.44±0.03 | -4.40±0.20 | 2.90±0.02 | 14 | 30 | 43 |
| **AFM B exciton** | 1.70±0.02 | -0.08±0.02 | 0.05±0.01 | -0.13±0.01 | 14 | 30 | 43 |